\documentclass[11pt]{article}
\usepackage{amssymb}
\usepackage{amsmath}

   \renewcommand{\theequation}{\thesection.\arabic{equation}}
   \csname @addtoreset\endcsname{equation}{section}
    \csname @addtoreset\endcsname{figure}{section}
    \csname @addtoreset\endcsname{table}{section}

\newcounter{thanksnum}
\def\thanksnumber#1
{\setcounter{thanksnum}{\value{footnote}}\setcounter{footnote}{#1}%
                     \addtocounter{footnote}{-1}\footnotemark
                     \setcounter{footnote}{\value{thanksnum}}}
\def\newtheoremz#1{\@ifnextchar[{\@othmz{#1}}{\@nthmz{#1}}}

\def\@nthmz#1#2{%
\@ifnextchar[{\@xnthmz{#1}{#2}}{\@ynthmz{#1}{#2}}}

\def\@xnthmz#1#2[#3]{\expandafter\@ifdefinable\csname #1\endcsname
{\@definecounter{#1}\@addtoreset{#1}{#3}%
\expandafter\xdef\csname the#1\endcsname{\expandafter\noexpand
  \csname the#3\endcsname \@thmcountersepz \@thmcounterz{#1}}%
\global\@namedef{#1}{\@thmz{#1}{#2}}\global\@namedef{end#1}{\@endtheoremz}}}

\def\@ynthmz#1#2{\expandafter\@ifdefinable\csname #1\endcsname
{\@definecounter{#1}%
\expandafter\xdef\csname the#1\endcsname{\@thmcounterz{#1}}%
\global\@namedef{#1}{\@thm{#1}{#2}}\global\@namedef{end#1}{\@endtheoremz}}}

\def\@othmz#1[#2]#3{\expandafter\@ifdefinable\csname #1\endcsname
  {\global\@namedef{the#1}{\@nameuse{the#2}}%
\global\@namedef{#1}{\@thmz{#2}{#3}}%
\global\@namedef{end#1}{\@endtheoremz}}}

\def\@thmz#1#2{\refstepcounter
    {#1}\@ifnextchar[{\@ythmz{#1}{#2}}{\@xthmz{#1}{#2}}}

\def\@xthmz#1#2{\@begintheoremz{#2}{\csname the#1\endcsname}\ignorespaces}
\def\@ythmz#1#2[#3]{\@opargbegintheoremz{#2}{\csname
       the#1\endcsname}{#3}\ignorespaces}

\def\@thmcounterz#1{\noexpand\arabic{#1}}
\def\@thmcountersepz{.}
\def\@begintheoremz#1#2{ \trivlist \item[\hskip \labelsep{\bf #1\ #2}]}
\def\@opargbegintheoremz#1#2#3{ \trivlist
      \item[\hskip \labelsep{\bf #1\ #2\ (#3)}]}
\def\@endtheoremz{\endtrivlist}

\newtheorem{theorem}{Theorem}[section]
\newtheorem{lemma}{Lemma}[section]

\newtheorem{definition}{Definition}[section]
\newtheorem{remark}{Remark}[section]

\def\e{\varepsilon}

\def\defi{\stackrel{{\scriptscriptstyle \Delta}}{=}}

\def\a{\alpha}
\def\d{\delta}
\def\o{\omega}
\def\O{\Omega}

\def\Y{{\cal Y}}
\def\F{{\cal F}}
\def\w{\widehat}
\def\Ind{{\,\rm Ind\,}}
\def\Ind{{\mathbb{I}}}

\def\esssup{\mathop{\rm ess\, sup}}
\def\const{{\rm const\,}}

\def\R{{\bf R}}
\def\E{{\bf E}}
\def\P{{\bf P}}

\def\H{{\cal H}}

\def\b{\beta}
\def\s{\delta}
\def\g{\gamma}

\def\ww{\widetilde}
\def\X{{\cal X}}
\def\t{\theta}
\def\oo{\bar}
\def\s{\sigma}
\def\p{\partial}

\def\U{{\cal U}}

\def\u{u}

\newcommand{\be}{\begin{equation}}
\newcommand{\ee}{\end{equation}}
\newcommand{\bd}{\begin{displaymath}}
\newcommand{\ed}{\end{displaymath}}
\newcommand{\ba}{\begin{array}{ll}}
\newcommand{\ea}{\end{array}}
\newcommand{\baa}{\begin{eqnarray}}
\newcommand{\eaa}{\end{eqnarray}}
\newcommand{\baaa}{\begin{eqnarray*}}
\newcommand{\eaaa}{\end{eqnarray*}}
\font\sm=cmr10
\def\oo{\bar}
\def\f{f}

\def\U{{\cal U}}
\def\f{{\rm f}}\def\gg{{\rm g}}
\date{\  Submitted: December 6, 2010; revised: October 14, 2011}
\title{
Controlled options: derivatives with added flexibility  % and their pricing %
}
\author{
Nikolai Dokuchaev\\
 {\sm Department of Mathematics \& Statistics, Curtin
University,}\\ {\sm  GPO Box U1987, Perth, 6845 Western
Australia}%\\{\sm  email N.Dokuchaev@curtin.edu.au}
}
\begin{document}
\maketitle
\begin{abstract}
The paper introduces  a limit version of multiple stopping options
such that the holder selects dynamically a weight function that
control the distribution of the payments (benefits) over time. In
applications for
 commodities  and energy trading, a control process
can represent the quantity  that can be purchased by a fixed price
at current time. In another example, the control represents the
 weight of the integral in a modification of the Asian option.
The pricing for these options requires to solve  a stochastic
control
 problem. Some existence results and pricing rules are obtained via modifications of parabolic
 Bellman equations.
\
%\\ {\bf 2000 Mathematics Subject Classification: 91B70,  %Stochastic models
%93E20% Optimal stochastic control
%D52, % Incomplete Markets
%D84, % Expectations; Speculations ,
%G11 % Portfolio Choice; Investment Decisions
%of other theories (stochastic control, calculus of variations, PDE,SPDE, dynamical systems)
%93E20,      % Optimal stochastic control
 % 91G10       %Portfolio theory
\end{abstract}
%{\it Short running head}: {\sm }

\section{Introduction}
There are many different types of options and financial derivatives:
European, American, Asian, Bermudian, Israeli, Russian, Parisian
options, etc. (see, e.g., Briys {\it et al} (1998)). Pricing of
exotic options require special methods; see, e.g.,
 Kifer (2000), Kyprianou
(2004), Kramkov {\it et al} (1994), Meinshausen and Hambly (2004),
Peskir (2005), Bender and Schoenmakers (2006), Bender (2011),
Carmona and Dayanik (2006), Carmona and Touzi (2006), Dai and Kwok
(2008), Dokuchaev (2009).
\par
Typically, new type of options are developed with the purpose to
offer some additional flexibility for an option holder. In
particular, multiple exercise options are used in energy trading;
allow a holder to distribute the purchase of energy by a fixed price
over a set of time moments; see, e.g.,  Meinshausen and Hambly
(2004), Bender and Schoenmakers (2006), Bender (2011), Carmona and
Dayanik (2006), Carmona and Touzi (2006), Dai and Kwok (2008),
Dokuchaev (2009), and Bender (2011). The paper suggests a next step
in this direction.  We develop a family of options that allows the
holder to select dynamically continuous time processes that control
the payoff. We call the new options {\em controlled
 options}.
 The  control processes  are assumed to be adapted to the current
 flow of information. More precisely,
the holders of the new options select dynamically the weight
functions that control the distribution of the payments (benefits)
over time.  There is  a similarity with passport options introduced
in Hyer {\em et al} (1997) as a generalization of the American
option (see also Delbaen  and Yor (2002), Kampen (2008), Nagayama
(1999)). The passport options
 allow the holder to select investment strategies for an account; the writer
 guarantees protection from the losses.  The difference
 with the control options introduced in this paper is that the holder of passport options selects
 portfolio strategy.
\par
 The controlled options may have applications in commodities and energy
 trading.  For instance, control process  $u(t)$ may represent the
 weight of the integral in a modification of the Asian option. In another example, a
non-negative control process $u(t)$ can represent the amount
 of some commodity that can be purchased by a certain given price at time $t\in[0,T]$,
 where $T$ is the terminal time, given that $\int_0^Tu(t)dt=1$.
 This is a limit case of multi-exercise option studied in Bender and Schoenmakers (2006) and Bender (2011), where
 the distribution of exercise times approaching a continuous
 distribution.  Therefore, controlled options can be used also as an auxiliary tool to study
these multi-exercise options.  In some cases, analysis of these
controlled option is more straightforward since optimal
multi-stopping is actually excluded; it is replaced by more standard
stochastic control problem.
 These and other examples of controlled options
 are studied below.
 It is shown that pricing for these options requires solution of a stochastic control problem
 rather than optimal stopping problem. Some existence results pricing rules are obtained in Markov diffusion setting based on
dynamic programming and various modifications of degenerate
parabolic Bellman equations (Hamilton-Jacobi-Bellman (HJB)
equation).

The paper is organized as follows. In Section \ref{SecG}, two
classes of controlled options are introduced: (i) options where the
adapted weight $u(t)$ is selected such that $\int_0^1u(t)dt=1$, and
(ii) options where the weight $u(t)$ does not restriction on its
cumulate, and where the payoff is defined by the normalized weight
$v(t)=\left(\int_0^Tu(s)ds\right)^{-1}$ which is not adapted. Some
motivation for this setting is given.  In Section \ref{SecM}, the
market model is introduced. In Section \ref{secFP}, the general
martingale pricing formula is given. In Section \ref{secC}, the
pricing is discussed for the case (i). In Section \ref{secNonC}, the
pricing is discussed for the case (ii). The proofs are given in
Appendix.
\section{Controlled options:
definition and examples}\label{SecG} Consider a  risky asset (stock,
commodity, a unit of energy) with the price $S(t)$, where
$t\in[0,T]$, for a given $T>0$.  Consider an option with the payoff
\baa F_u=\Phi(u(\cdot),S(\cdot)). \label{Phi}\eaa
 This payoff depends on a
control process $u(\cdot)$ that is selected by an option holder from
a certain class of admissible controls $\U$.  The mapping
$\Phi:\U\times {\cal S}\to\R$ is given; ${\cal S}$ is the set of
paths of $S(t)$. All processes from $\U$ has to be adapted to the
current information flow, i.e., adapted to some filtration $\F_t$
that describes this information flow.
\par
We call the corresponding options controlled options. Clearly, an
American option is a special case of controlled options, where the
exercise time is selected. Some new examples of controlled options
are suggested and discussed below.
\par
For simplicity, we assume that all options give the right on the
corresponding  payoff of the amount $F_u$ in cash rather than the
right to buy or sell stock or  commodities.
 \subsection*{Options with adapted weight with fixed cumulated integral}
%\subsubsection*{Setting when admissible weights  $u(t)$ are from the class of adapted densities on $[0,T]$}
 Consider a  risky asset  with the price $S(t)$. Let $T>0$ be given, and let $g:\R\to\R$ and
 $f:\R\times[0,T]\to\R$ be some functions.
Consider an option with the payoff at time $T$ \baa
F_u=g\left(\int_0^Tu(t)f(S(t),t)dt\right), \label{A3}\eaa
 Here $u(t)$ is the control process that is selected
by the option holder. The process $u(t)$ has to be adapted to the
filtration $\F_t$  describing the information flow. In addition, it
has to be selected such that \baaa \int_0^Tu(t)dt= 1.
\label{M4}\eaaa
%\begin{remark}{\rm
\par
A possible modification  is the option with the payoff \baaa
F_u=F_u=\int_0^Tu(t)f(S(t),t)dt+\left(1-\int_0^Tu(t)
dt\right)f(S(T),T).
 \eaaa
 In this case, the unused $u(t)$ are accumulated and used at the terminal time.
\par
Let us consider some examples of possible selection of $f$ and $g$.

We denote $x^+\defi \max(0,x)$.

Important special cases are the options with $g(x)=x$,
$g(x)=(x-k)^+$, $g(x)=(K-x)^+$, $g(x)=\min(M, x)$, where $M>0$ is
the cap for benefits, and with \baa f(x,t)=x,\quad f(x,t)=(x-K)^+,
\quad f(x,t)=(K-x)^+, \label{putcall}\eaa or \baa
f(x,t)=e^{r(T-t)}(x-K)^+,\quad f(x,t)=e^{r(T-t)}(K-x)^+,
\label{rputcall}\eaa where $K>0$ is given and where $r>0$ is the
risk-free rate.

\par Options
(\ref{putcall}) correspond to the case when the payments are made at
current time $t\in[0,T]$, and options (\ref{rputcall}) correspond to
the case when the payment is made at terminal time $T$. This model
takes into account accumulation of interest up to time $T$ on any
payoff.
\par
 The option with
payoff (\ref{A3}) with $f(x,t)\equiv x$ represents a generalization
of Asian option where the weight $u(t)$ is selected by the holder.
\par
 The option with
payoff (\ref{A3}) with $g(x)\equiv x$  represents a limit version of
the multi-exercise options, when the distribution of exercise time
approaches a continuous distribution. An additional  restriction on
$|u(t)|\le const$ would represent the continuous analog  of the
requirement for multi-exercise options that exercise times must be
on some distance from each other. For an analog of the model without
this condition, strategies that may approach delta-functions.
\par
These options can be used, for instance, for energy trading with
$u(t)$ representing the quantity of energy purchased at time $t$ for
the fixed price $K$ when the market price is above $K$. In this
case, the option represents a modification of the multi-exercise
call option with continuously distributed payoff time. For this
model, the total amount of energy that can be purchased is limited
per option. Therefore, the option holder may prefer to postpone the
purchase if she expects better opportunities in future.

\index{ The total amount $\int_0^Tu(t)dt$ of energy that can be
purchased is not fixed and can be selected by the option holder, but
the price of one unit is predetermined. Therefore, the option holder
can  buy larger amount if the price is low.
 A modification  with $f(S(t),t)=S(t)$
can be used to describe a service contract with fixed price for a
unit and with the number of service units selected by the contact
holder. In this model, $u(t)$ describes the quantity of units
requested at given time, and $S(t)$ represents the current
stochastic market price of the service.  The pricing of this
modification of the option means selecting the fair predetermined
price for service per one unit. }
\subsection{Option with non-adapted normalized weight} A possible modification of the option described above is the option
with the payoff  at time $T$ \baa
F_u=g\left(\int_0^Tv(t)f(S(t),t)dt\right),\label{A1}\eaa where
$g:\R\to \R$ and $f:\R\times [0,T]\to \R$ are given functions, the
process $v(t)$ is such that  \baaa \int_0^Tv(t)dt=1. \eaaa  The
difference with option (\ref{A3}) is that the process $v(t)$ is not
assumed to be adapted to the filtration  $\F_t$ generated by the
current information flow. It is formed as
 \baa v(t)= \frac{u(t)}{\int_0^Tu(s)ds},
\label{A1v}\eaa where the  process $u(t)$ is selected by the option
holder dynamically, using the current flow of information, i.e., it
has to be adapted to the filtration $\F_t$ describing this flow. The
process $u(t)$ can be called weight process, and $v(t)$ can be
called normalized wealth process.
\par
This setting means that the option holder keeps the writer informed
about her current selection  of the value of $u(t)$, and these
choices are recorded; the payoff occurs ate terminal time $T$.

\par
We don't exclude the case when $d_0=0$ and $u(t)|_{t\in[0,T)}= 0$.
In this case, the payoff  can be set by different ways. A possible
way is to define the  payoff for $u(t)\equiv 0$ as $
F_u=g\left(\frac{1}{T}\int_0^TS(t)dt\right),$ i.e., as the limit of
the payoff (\ref{A1}) for $u(t)\equiv \e$ as $\e\to 0$. Another
possible selection of the  payoff for $u(t)\equiv 0$ is $
F_u=g\left(f(S(T),T)\right),\index{delta0}$ i.e., as the limit of
the payoff (\ref{A1}) for $u(t)\equiv \e\Ind_{\{t\ge T-\e\}}$ as
$\e\to 0$. In this case, $v(t)$ can be interpreted as the delta
function with the mass concentrated at $t=T$.
\par
These options can be useful generalizations of Asian options.
Consider, for instance, a customer who consumes time variable and
random quantity $u(t)$ of energy per time period $(t,t+dt)$, with
the price $S(t)$ for a unit. The cumulated number of units consumed
up to time $T$ is $\oo u=\int_0^Tu(t)dt$; it is  unknown at times
$t<T$. To hedge against the price rise, the customer would purchase
a portfolio of $M$ call options; each option  gives the right to
purchase one energy unit for the price $K$. To minimize the impact
of price fluctuations, the Asian options are commonly used. These
options can be described as the options with the payoff $(\oo
S-K)^+$, where $\oo S= T^{-1}\int_0^TS(t)dt$. For accounting and tax
purposes, the average price of energy for a particular customer has
to be calculated as $\oo S_u=\oo u^{-1}\int_0^Tu(t)S(t)dt$ rather
than $\oo S$. Therefore, more certainty in financial and tax
situation can be achieved if one uses the portfolio of $M$ options
with payoff $(\oo S_u-K)^+$ that is defined by the consumption of
the particular customer. This is a special case of option
(\ref{A1}). Since $\oo u$ is random and unknown,  options (\ref{A3})
cannot be used for this model.

\subsection*{On impact of fixing the cumulated $u(t)$} It may appear
that options with payoffs (\ref{A1})  are equivalent to the related
options (\ref{A3}). However, the nature of control for these options
is different.

First, the selection of $u(t)$ is obviously more restricted for
options (\ref{A3})  than for options  (\ref{A1}): the option holder
have to obey the restrictions on the total amount of cumulated
$u(t)$.  Second, these two types of the options have different
opportunities with respect to possibility to correct past decisions.
Consider a model where the option holder selects $u(t)$ with the
purpose to maximize the payoff $F_u$.  For the holders of options
(\ref{A1}), it is possible to smooth the effect of unfortunate
decisions made at previous times by selecting larger $u(t)$ at
future times.  In addition,  the relative weight of the past good
decisions can be enlarged via selecting small current $u(t)$. This
opportunity is absent
for options (\ref{A3}). %}\end{remark}
\index{Note that this additional flexibility still does not allow to
compensate missing good opportunities.}
\section{Market model}\label{SecM}
 We investigate pricing of the options described above for the
 simplest case of Black-Scholes model, i.e,
 for a complete continuous time diffusion market model with constant volatility.  We consider the model of a
securities market  consisting of a risk free bond or bank account
with the price $B(t)$ and
 a risky stock with the price $S(t)$, $t\in[0,T]$, where $T>0$ be given terminal time.
\par
Ut to the end of this paper, we assume that the prices of the stocks
evolves as
 \be \label{S}
dS(t)=S(t)\left(a(t) dt+\s dw(t)\right), \ee where $a(t)$ is an
appreciation rate, $\s>0$ is a volatility coefficient.
\par
In (\ref{S}),
 $w(\cdot)$ is a standard Wiener process on a given
standard probability space $(\O,\F,\P)$, where $\O=\{\o\}$ is a set
of elementary events, $\F$ is a complete $\s$-algebra of events, and
$\P$ is a probability measure.
\par
  The
price of the bond evolves as
\begin{equation}
\label{B} B(t)=e^{rt}B(0).
\end{equation}
We assume that $\s>0$, $r\ge 0$, $B(0)>0$, and $S(0)>0$, are given
constants.
\par
 Let $\F_t$
be the filtration generated by $w(t)$. For simplicity, we assume
that $a(t)$ is a bounded process progressively measurable with
respect to $\F_t$. In this case, $\F_t$ is also the filtration
generated by $S(t)$.
 \par
Let $\P_*$ be the probability measure such that the process
$e^{-rt}S(t)$ is a martingale under $\P_*$ on $[0,T]$.  By the
assumptions for $(a,\s,r)$, this measure exists and it is unique.
Let $\E_*$ be the corresponding expectation. Under the risk neutral
measure $\P_*$,  $$ w_*(t)\defi w(t)+\int_0^t\s^{-1}[a(s)-r]ds$$ is
a Wiener process, and the process $\ww S(t)$  is a martingale, since
$a(t)dt+\s dw(t)=\s d w_*(t)$ and $d\ww S(t)=\s dw_*(t)$.
 \subsection*{Admissible portfolio strategies}
Let $X(0)>0$ be the initial wealth at time $t=0$ and let $X(t)$ be
the wealth at time $t>0$. We assume that the wealth $X(t)$ at time
$t\in [0,T]$ is
\begin{equation}
\label{X} X(t)=\b(t)B(t)+\g(t)S(t).
\end{equation} Here $\b(t)$ is
the quantity of the bond portfolio, $\g(t)$ is the quantity of the
stock  portfolio, $t\ge 0$. The pair $(\b(\cdot), \g(\cdot))$
describes the state of the bond-stocks securities portfolio at time
$t$. Each of  these pairs is  called a strategy.
\par
The process $ \ww X(t)\defi e^{-rt}X(t)$ is said to be the
discounted wealth, and the process $ \ww S(t)\defi e^{-rt}S(t)$ is
said to be the discounted stock price.
\begin{definition}
\label{adm} A pair $(\b(\cdot),\g(\cdot))$  is said to be an
admissible strategy if $\b(t)$ and $\g(t)$ are  random processes
which are progressively measurable with respect to the filtration
$\F_t$ and such that there exists a sequence of Markov times
$\{T_k\}_{k=1}^{+\infty}$ with respect to the filtration $\F_t$ such
that $T_k\to T-0$ a.s. and $$ \E\int_0^{T_k}\left(\b(t)^2B(t)^2+
S(t)^2\g(t)^2\right)dt<+\infty\qquad \forall k=1,2,... $$
\end{definition}
\begin{definition}
A pair $(\b(\cdot),\g(\cdot))$  is said to be an admissible
self-financing strategy, if
\begin{equation}
\label{self} dX(t)=\b(t)dB(t)+\g(t)dS(t).
\end{equation}
\end{definition}
It is well known that (\ref{self}) is equivalent to \be d\ww
X(t)=\g(t)d\ww S(t). \label{wwXS}\ee  It follows that $\ww X(t)$ is
a martingale with respect to the probability measure $\P_*$.
\par
\index{We assume that all Markov times mentioned below are Markov
times with respect to the filtration $\F_t$. \par
 In addition,
we assume that the wealth and strategies are defined for Markov
random initial times $\t$ such that $\t\in[0,T]$. All definitions
can be easily rewritten for this case.} \par Let $X(0)$ be an
initial wealth, and let $\ww X(t)$ be the discounted wealth
generated by an admissible self-financing strategy
$(\b(\cdot),\g(\cdot))$. For any Markov time $\tau$ such that
$\tau\in[0,T]$, we have  $$ \E_*\ww
X(\tau)=X(0)+\E_*\int_0^\tau\g(t)d\ww
S(t)=X(0)+\E_*\int_0^\tau\g(t)\ww S(t)^{-1}d w_*(t)=X(0). $$
\par
\index{Similarly, let $X(\t)$ be an initial wealth at Markov initial
time $\t$ such that $\t\in[0,T]$, and let $\ww X(t)$ be the
discounted wealth generated by an admissible self-financing strategy
$(\b(\cdot),\g(\cdot))$. For any Markov time $\tau$ such that
$\tau\in[\t,T]$, we have  \baaa \E_*\{\ww X(\tau)|\F_\t\}=\ww
X(\t)+\E_*\Bigl\{\int_\t^\tau\g(t)d\ww S(t)\Bigl|\F_\t\Bigr\} =\ww
X(\t)+\E_*\Bigl\{\int_\t^\tau\g(t)\ww S(t)^{-1}d
w_*(t)\Bigl|\F_\t\Bigr\}\\=\ww X(\t). \eaaa}
\par
\index{ We shall use the extended definition of the strategy
assuming that a strategy may include buying and selling the bonds,
the stock, and the options. All  transactions must be
self-financing; they represent redistribution on the wealth between
different assets, and there is  income from external sources. Short
selling is allowed for the bonds and the stocks and is not allowed
for the options. For instance, a trader may borrow amount of money
$x$ to buy $k$ American options at time $t$ with payoff
$F(S(s))|_{s\in[t,T]}$, then his/her total wealth at exercise time
$s$ is $kF(S(s))-e^{r(s-t)}x$. We assume that the standard
definition of self-financing strategy is extended for these
strategies.}
\section{The fair price of a general controlled option}\label{secFP}
Let us consider a controlled option  (\ref{Phi}) with the payoff  $
F_u=\Phi(u(\cdot),S(\cdot))$,   where $\Phi:\U\times C(0,T)$ is a
measurable mapping such that $\sup_{u\in\U}\E_* |F_u|<+\infty$. Here
$\U$ is the set of all admissible controls $u(\cdot)$. All examples
considered above are covered by this general setting.
\begin{definition}
The fair price of an option is the price $c$ such that
\begin{itemize}
\item
The option writer cannot fulfill option obligations at terminal time
$T$ using the wealth raised from the initial wealth $X(0)<c$ with
self-financing strategies. \index{The option writer can fulfill
option obligations using the wealth raised from the initial $X(0)=c$
using self-financing strategies.}
\item A rational option buyer would't buy an option for a higher
price than $c$.
\end{itemize}
\end{definition}
The following theorem is formulated for the case of constant $r$.
However, this theorem  holds for any model where the risk-neutral
measure $\P_*$ exists and is unique; the extension on the case of
time variable $r=r(t)$ is straightforward.
\begin{theorem}\label{ThM}
The fair price $c_F$ of an option with the payoff $F_u$ is \baaa
c_F=e^{-rT}\sup_{u\in \U}\E_* F_u. \eaaa
\end{theorem}

Proofs are given in the Appendix.
\section{Pricing of options with adapted weight}\label{secC}
Consider an option with payoff \baa
F_u=g\left(\int_0^Tu(t)f(S(t),t)dt\right), \label{Foptg} \eaa
 where $f(x,t):(0,+\infty)\times[0,T]\to \R$ and $g(x):(0,+\infty)\to \R$
are given continuous non-negative  functions such that
$|f(x,t)|+|g(x)|\le \const(|x|+1)$ and  $|\p f(x,t)/\p x|+| dg(x)/d
x|\le \const$. In addition, we assume that the function $g(x)$ is
non-decreasing.
\par
The function $u(t)$ is the control process that is selected by the
option holder.

We assume that  $S(t)$ and $\F_t$ are such as described in section
\ref{SecM}.

Let $\U$ be the class of  processes $u(t)$ consisting of the
processes that are adapted to the filtration $\F_t$ and such that
\baa u(t)\in[d_0,d_1], \label{M21}\eaa where $0\le d_0< d_1<
+\infty$.
\par
We consider  the class $\U_1$ of admissible processes $u(t)$
consisting of the processes $u\in\U$  such that \baa
\int_0^Tu(t)dt=1. \label{FM21}\eaa To ensure that the set of
admissible strategies in non-empty, we assume that $d_0T<1$.
\par
By Theorem \ref{ThM}, the fair price of this option is
 \baa c_F=e^{-rT}\sup_{u(\cdot)\in U_1} \E_*F_u. \label{cum}\eaa
%\subsection{Existence of optimal control}
\begin{lemma}\label{lemmaEc} Assume that the function $g$ is concave on $(0,+\infty)$.
In this case, an optimal control for problem (\ref{cum}) exists in
$\U_1$.
\end{lemma}
\subsection{Pricing via dynamic programming}
It follows from the definitions that the price $c_F$ for this option
can be found via solution of optimal stopping  problem \baa
\hbox{Maximize}\quad &&\E_*g(x(\tau)) \quad\hbox{over}\quad
u(\cdot)\in \U,\nonumber \\
\hbox{\, subject to}\quad  &&d x(t)=u(t)f(S(t),t)dt,
\nonumber\\
&&d y(t)=u(t)dt,
\nonumber\\
&&dS(t)=r S(t)dt+\s S(t) dw_*(t), \label{FAsO}\eaa  where $
\tau=T\land \inf\{t\in[0,T]:\ y(t)\ge 1\}.$ \index{tau} In this
case, $c_F=e^{-rT}\sup_{u(\cdot)\in \U} \E_*g(x(\tau))$  given that
$x(0)=0,\ y(0)=0,\ S(0)=S_0$.
\par
Alternatively, the price $c_F$ for this option can be found via
solution of optimal stopping stochastic control problem \baa
\hbox{Maximize}\quad &&\E_*g(x(T)) \quad\hbox{over}\quad
u(\cdot)\in \U,\nonumber\\
\hbox{\, subject to}\quad  &&d x(t)=\Ind_{\{y(t)<1
\}}u(t)f(S(t),t)dt,
\nonumber\\
&&d y(t)=u(t)dt,
\nonumber\\
&&dS(t)=r S(t)dt+\s S(t) dw_*(t).\label{cumS}\eaa In this case,
$c_F=e^{-rT}\sup_{u(\cdot)\in \U} \E_*g(x(T))$  given that $x(0)=0,\
y(0)=0,\ S(0)=S_0$.
\par  Problem (\ref{FAsO}) and (\ref{cumS}) are    such that   the  matrix
of the diffusion coefficients for the state process is degenerate.
In addition, problem (\ref{FAsO}) involves first exit from a  domain
with a boundary, This makes it difficult to use classical methods of
solution. Hence it will be more convenient to use (\ref{cumS}) that
does not feature a boundary and first exit time.
\par
The state equation for problem (\ref{cumS})  has discontinuous drift
coefficient for $x(t)$. To remove this feature, we approximate the
problem as the following.\index{ For $\e>0$, set
$\f_\e(x,t)=\min(f(x,t),\e^{-1})$, $h_\e(y)=1$, $y<1-\e$,
$h_\e(y)=\e^{-1}(1-y)y$, $y\in(1-\e,1)$, $h_\e(y)=0$, $y>1$.}
\par
\def\ff{{\bf f}}\def\ggg{{\bf g}}
Let functions $\phi_\e(x,t):(0,+\infty)\to\R$  be such as described
in Section \ref{secNonC}, $\e> 0$. Let functions
$\ggg_\e(x):\R\to\R$ and $\xi_\e(y):\R\to[0,1]$ be selected such
that the following holds.
\begin{enumerate}
\item The functions $\xi_\e$ are  non-increasing continuously
differentiable and such that $\xi_\e(y)=1$ for $y<T-\e$, and
$\xi_\e(y)=0$ for $y>1-\e+\e^2$.
\item
 The functions
$\ggg_\e(x)$ are bounded and twice differentiable. The corresponding
derivatives are bounded, and \vspace{-0.5cm}
 \baaa   &&\ggg_\e(x)\to g(x)\quad \hbox{as}\quad
\e\to 0,\qquad \ggg_\e(x)\le g(x) \quad \hbox{for all}\quad x.
\eaaa
\end{enumerate}
\par
Let $\ff_\e(u,x,y,s)=u\xi_\e(y)\phi_{\e}(x,t)$. \par Consider the
stochastic control following problem: \baa \hbox{Maximize}\quad
&&\E_*\ggg_\e(x(T))
\quad\hbox{over}\quad u(\cdot)\in \U,\nonumber\\
\hbox{\, subject to}\quad  &&d x(t)=\ff_\e(y(t),u(t),S(t),t)dt,
\nonumber\\
&&d y(t)=u(t)dt,
\nonumber\\
&&dS(t)=r S(t)dt+\s S(t) dw_*(t). \label{FAsO2}\eaa Consider the
corresponding value function \baa
 J_\e(x,y,s,t)
 \defi
\sup_{u(\cdot)\in\U}\E_*\Bigl\{\ggg_\e\left(x_\e(T)\right)\Bigl|
x(t)=x,\, y(t)=y,\,S(t)=s\Bigr\}. \label{J1}\eaa
\par
Let $D=\R\times \R\times \R\times [0,T]$. Let $\X$ be the class of
functions $v(x,y,z,t):D\to \R$ such that $v$ is continuous and there
exists $c>0$ such that $v(x,y,z,t)\le c(|x|+|y|+|z|+1)$ for all
$(x,y,z,t)\in D$. Let $\X_1$ be the class of functions $v\in \X$
such $v'_x$, $v'_y$, and $v'_z$ belong to $\X$. Let $\X_2$ be the
class of functions $v\in \X_1$ such that $v'_t$ and $v''_{zz}$
belong to $\X$.
\begin{theorem}\label{ThBEc} \begin{enumerate}
\item The option price can be found as
\baa c_F=\lim_{\e\to 0} e^{-rT}J_\e(0,0,S(0),0).\label{clim2}\eaa
\item
The value function $J=J_\e$ satisfies the Bellman equation
 \baa
&&J_t+  \max_{u\in [d_0,d_1]}  \{ J_x' \ff_\e + J_y' u\}+ J_s'
rs+{\scriptstyle \frac{1}{2}} J_{ss}''\s^2s^2 = 0,\nonumber
\\
&&J(x,y,s,T)=\ggg_\e(x). \label{BelEq}\eaa   The Bellman equation
has unique solution in the class of functions $J=J_\e(x,y,s,t)$ such
that $J_\e(x,y,s,t)=V_\e(x,y,\log s,t)$ for some function
$V_\e\in\X_2$. The Bellman equation holds as an equality that is
satisfied for a.e.
$(x,y,s,t)\in\R\times\R\times(0,+\infty)\times[0,T]$.
\end{enumerate}
\end{theorem}
\subsection{Case of linear $g$} The
dimension of the Bellman equation can be reduced for the case when
$g(x)\equiv x$. In this case, the option price $c_F$   can be found
via solution of optimal stopping  problem \baa \hbox{Maximize}\quad
&&\E_*\int_0^\tau u(t)f(S(t),t)dt, \quad\hbox{over}\quad
u(\cdot)\in \U,\nonumber\\
\hbox{\, subject to}\quad  &&d y(t)=u(t)dt,
\nonumber\\
&&dS(t)=r S(t)dt+\s S(t) dw_*(t), \label{FFAsO}\eaa where $
\tau=T\land \inf\{t\in[0,T]:\  y(t)\ge 1\}$. In this case,
$c_F=e^{-rT}\sup_{u(\cdot)\in \U} \E_*\int_0^{\tau}u(t)f(S(t),)dt$
given that $y(0)=0,\ S(0)=S_0$.  Consider the corresponding value
function \baa {\oo J(y,s,t)}
 \defi
\sup_{u(\cdot)\in\U}\E_*\Bigl\{ \int_t^{\tau_u^{x,t}}
u(s)f(S(s),s)ds\Bigl| y(t)=y,\,S(t)=s\Bigr\}. \label{J2}\eaa  Here
\baaa \tau_u^{y,s}=T\land \inf\{\t\in[t,T]:\ y+\int_s^{\t}u(q)dq\ge
1\}. \label{tau2}\eaaa The option price is $ c_F=e^{-rT}{\oo
J}(0,S(0),0)$. The  Bellman equation satisfied formally by
 ${\oo J}$ is
 \baa
&&{\oo J}_t + \max_{u\in [d_0,d_1)}  \{ {\oo J}_y' u +uf(s,t)\}+{\oo
J}_s' rs+{\scriptstyle \frac{1}{2}} {\oo
J}_{ss}''\s^2s^2=0,\nonumber
\\
&&{\oo J}(1,s,T)=0, \qquad {\oo J}(y,s,T)=0. \label{FFBelEq}\eaa The
Bellman equation holds for $x>0$, $y<1$, $s>0$, $t<T$.  However, to
derive this equation and prove Verification Theorem, one have to
overcome again some technical difficulties arising from the presence
of boundary and from the fact that the diffusion in the state
equation is degenerate. Instead,
 we suggest to use an alternative  stochastic control
problem \baaa \hbox{Maximize}\quad &&\E_*\int_0^T \Ind_{\{y(t)\le
1\}}u(t)f(S(t),t)dt, \quad\hbox{over}\quad
u(\cdot)\in \U,\nonumber\\
\hbox{\, subject to}\quad  &&d y(t)=u(t)dt,
\nonumber\\
&&dS(t)=r S(t)dt+\s S(t) dw_*(t). \label{FFAsO2}\eaaa This  problem
without does not involve first exit time.  The Bellman equation
satisfied formally by its value  function $J=J(y,s,t)$ is \baaa
&&J_t+ \max_{u\in [d_0,d_1]}  \{ J_x' uf  + J_y' u+\Ind_{\{y\le
1\}}uf\} + J_s'rs+{\scriptstyle \frac{1}{2}} J_{ss}''\s^2s^2 =
0,\nonumber
\\
&&J(y,s,T)=0. \label{BelEq2}\eaaa The question is degenerate again,
so we will use an equation with more regular coefficients as an
approximation.
\par
 Let functions $\ff_\e$ be such as described above.
Consider stochastic control problem \baa \hbox{Maximize}\quad
&&\E_*\int_0^T \ff_\e(u(t),y(t),S(t),t)dt, \quad\hbox{over}\quad
u(\cdot)\in \U,\nonumber\\
\hbox{\, subject to}\quad  &&d y(t)=u(t)dt,
\nonumber\\
&&dS(t)=r S(t)dt+\s S(t) dw_*(t). \label{FFAsO22}\eaa Consider the
corresponding value function \baa
 J_\e(y,s,t)
 \defi
\sup_{u(\cdot)\in\U} \E_*\Bigl\{ \int_t^T
\ff_\e(u(t),y(t),S(t))dt\Bigl| y(t)=y,\,S(t)=s\Bigr\}.
\label{Jcl}\eaa
\par Let $D'=\R\times \R\times [0,T]$. Let $\Y$ be
the class of functions $v(y,z,t):D'\to \R$ such that $v$ is
continuous and there exists $c>0$ such that $v(y,z,t)\le
c(|y|+|z|+1)$ for all $(y,z,t)\in D'$. Let $\Y_1$ be the class of
functions $v\in \Y$ such  $v'_y$ and $v'_z$ belong to $\Y$. Let
$\Y_2$ be the class of functions $v\in \Y_1$ such $v'_t$ and
$v''_{zz}$ both belong to $\Y$.

\begin{theorem}\label{ThBEl}
\begin{enumerate}
\item The option price can be found as
\baa c_F=\lim_{\e\to 0} e^{-rT}J_\e(0,S(0),0).\label{clim3}\eaa
\item
The value function $J=J_\e(y,s,t)$ for problem (\ref{FFAsO2})
satisfies the Bellman equation
 \baa
&&J_t +\max_{u\in [d_0,d_1]}  \{ J_y' u+\ff_\e\}+ J_s'
rs+{\scriptstyle \frac{1}{2}} J_{ss}''\s^2s^2 = 0,\nonumber
\\
&&J(y,s,T)=0. \label{BelEq3}\eaa The Bellman equation has unique
solution in the class of functions $J=J_\e(x,y,s,t)$ such that
$J_\e(y,s,t)=V_\e(y,\log s,t)$ for some function $V_\e\in\Y_2$. The
Bellman equation holds as an equality that is satisfied for a.e.
$(y,s,t)\in \R\times(0,+\infty)\times[0,T]$.
\end{enumerate}
\end{theorem}
%\end{document}
\subsection{Analog of Merton Theorem} In this section, we consider
again a risky asset with the price $S(t)$, where $t\in[0,T]$.
Consider an option with the payoff at time $T$  \baaa
F_u=\int_0^Tu(t)f(S(t),t)dt, \eaaa where
$f:(0,+\infty)\times[0,T]\to\R$ is a given function such that
$|f(x,t)|\le \const (1+|x|)$ and $f(x,t)\ge 0$. Here $u(t)$ is the
control process that is selected by the option holder. The set
$\U_1$ of admissible processes $u(t)$ consists of the processes that
are adapted to the current information flow (or to the filtration,
generated by $S(t)$ and such that \baaa u(t)\in[0,L],\quad
\int_0^Tu(t)dt\le 1, \eaaa where $L\in(0,+\infty)$ is given.
\par
If $TL\le 1$ then the optimal solution is  $u\equiv L$. Hence we
assume that $T>L^{-1}$.
\par
 This option represents the limit version of
multi-exercise options when the distribution of exercise times
approaching a continuous distribution. This model can be used, for
instance, for energy trading with $u(t)$ representing the quantity
of energy purchased at time $t$ for the fixed price $K$ when the
market price is above $K$. The total amount $\int_0^Tu(t)dt$ of
energy that can be purchased is limited per option.
\par
Merton theorem states that American and European options with the
same parameters have the same price and that early exercise is not
rational. The following theorem represents a extension of this
theorem on the case of controlled  options.
\begin{theorem}\label{ThAM}
Let $f(S(t),t)=e^{r(T-t)}h(S(t))$, where  the function $h(x)$ is
convex and non-linear in $x>0$, and such that at least one of the
following conditions holds:
\begin{itemize}
\item[(i)]
the function $\a^{-1} h(\a x)$ is non-decreasing in $\a\in(0,1]$; or
\item[(ii)] $r=0$.
\end{itemize}
Then $\sup_{u(\cdot)\in\U_1}\E_*F_u$
is achieved for the control process \baaa \w u(t)=\begin{cases}L,\quad t\ge T-1/L\\
0,\quad t<T-1/L,\end{cases} \eaaa and the price of the option is
\baaa e^{-rT}\E_*\int_0^T\w
u(t)f(S(t),t)dt=\frac{e^{-rT}}{L}\E_*\int_{T-1/L}^Tf(S(t),t)dt.
\eaaa
\end{theorem}

\begin{remark}{\rm
The function $h(x)=(x-K)^+$  is such that assumption (i) of Theorem
\ref{ThAM} are satisfied (this function corresponds to the call
option with continuously distributed payoff time). However,
assumptions (i) of Theorem \ref{ThAM} are not satisfied for
$h(x)=(K-x)^+$ that corresponds to the put option. The pricing for
this case with $r>0$ is an interesting problem. A solution could be
a useful approximation of the classical optimal stopping pricing
rule for American option: for the controlled option  with
restriction that $u(t)\in[0,L]$, the limit case when $L\to+\infty$
 will lead to a Stefan problem and optimal stopping. The approximate solution for finite $L$ may be easier to find since
it does not require optimal stopping and solution of Stefan problem.
  We leave it for future research. }\end{remark}
\section{Pricing for non-adapted normalized weight $v(t)$}\label{secNonC} Consider an option with payoff
(\ref{A1}),
 where functions  $f(x,t)$ and $g(x)$ have the same properties as in Section \ref{secC}.

The function $u(t)$ is the control process that is selected by the
option holder. The set of admissible processes $\U$ is  the set of
$\F_t$-adapted processes $u(t)$ that take values in $[d_0,d_1]$,
where $0\le d_0< d_1< +\infty$.
\par
We don't exclude the case when $d_0=0$ and $u(t)|_{t\in[0,T)}= 0$.
In this case, the payoff is assumed to be
 \baa
F_u=g\left(f(S(T),T)\right),\label{delta} \eaa i.e., as the limit as
$\e\to 0$ of the payoffs (\ref{A1}) defined for $u_\e(t)\equiv
d_1\Ind_{\{t>T-\e\}}$.
\par
\index{ For $\e>0$, let  $\Delta_\e(t)\equiv [d_0,d_1]$ if $d_0>0$.
If $d_0=0$, set $\Delta_\e(t)=[d_0,d_1]$ for $t\ge T-\e$ and
$\Delta_\e(t)=\{d_1\}$ for $t\ge T-\e$.
\par
 Let  $\U_\e=\U$
 if $d_0>0$, and let  $
 \U_\e=\{u(\cdot)\in \U:\quad u(t)\in\Delta_\e(t)
 \quad \forall t \quad\hbox{a.s.}
 \}$
 if $d_0=0$.}
\par
For $\e> 0$, let functions $\f_\e(u,x,t):(0,+\infty)\to\R$ and
$\gg_\e(x,y):\R^2\to\R$ be selected such that the following holds.
\begin{enumerate}
\index{item $\f_0(u,x,t)=uf(x,t)$, $\gg_0(x,y)=g(x/y)$ for all
$x,y,u,t$.}
\item
$\f_\e(u,x,t)=h_\e(u,t)\phi_\e(x,t)$, where $h_\e:[d_0,d_1]\times
[0,T]\to\R$ and $\phi_\e:(0,+\infty)\times[0,T]\to\R$ are measurable
functions with the following properties. \subitem(a) The functions
$\w \phi_\e(z,t)\defi \phi_\e(e^z,t)$ are bounded and continuously
differentiable in $(z,t)\in\R\times (0,T)$. The corresponding
derivatives are bounded. \subitem(b)
$h_\e(u,t)=u(1-\psi_\e(t))+d_1\psi_\e(t)$, where
$\psi_\e(t):\R\to[0,1]$ is a continuously differentiable
non-decreasing function  such that $\psi_\e(t)=0$ for $t<T-\e$,
$\psi_\e(t)=d_1$, $t>T-\e+\e^2$.
\item $\phi_\e(x,t)\le f(x,t)$, $\gg_\e(x,y)\le g(x/y)$ for all
$x,y\neq 0,t$.
\item
The functions $\gg_\e(x,y)$ are bounded and twice differentiable in
$(x,y)$. The corresponding derivatives are bounded. \item (a)
$\phi_\e(x,t)\to f(x,t)$ as $\e\to 0$ for all $x,t$.\\ (b) If $y\neq
0$ then $\gg_\e(x,y)\to g(x/y)\quad \hbox{as}\quad \e\to 0 $ for all
$x$.\\ (c)  If, for some $c\in\R$, we have that  $\e\to 0$, $y\to
0$, $x/y\to c$, then $\gg_\e(x,y)\to g(c)$.
\end{enumerate}
\index{Sdelat' postoyannye domain
 For $\e>0$, let $\f_\e(x,t)\equiv \min
(f(x,t),\e^{-1})$, $\gg_\e(x,y)=g(x/y)$ if $d_0>0$. If $d_0=0$, let
$\gg_\e(x,y)=g(x/y)$, $y>\e$, $\gg_\e(x,y)=g(x/\e)$, $y<\e$. }
\par
Consider optimal stochastic control problem \baa
\hbox{Maximize}\quad &&\E_*\gg_\e(x(T),y(T)) \quad\hbox{over}\quad
u(\cdot)\in \U,\nonumber\\
\hbox{\, subject to}\quad && d x(t)=\f_\e(u(t),S(t),t)dt,
\nonumber\\
&&d y(t)=h_\e(u(t),t)dt,
\nonumber\\
&&dS(t)=r S(t)dt+\s S(t) dw_*(t). \label{AsO}\eaa For $u\in\U$, set
\baaa {\cal J}_\e(u,x,y,s,t)\defi
\E_*\Bigl\{\gg_\e\left(x(T),y(T)\right)\Bigl| x(t)=x,\,
y(t)=y,\,S(t)=s\Bigr\}. \eaaa Consider the corresponding value
function \baa
 J_\e(x,y,s,t)
 \defi
\sup_{u(\cdot)\in\U}{\cal J}_\e(u,x,y,s,t). \label{J}\eaa
\par
Let $\X_2$ be the space introduced in Section
\ref{secC}.
\begin{theorem}\label{ThBE} \begin{enumerate}
\item The option price can be found as
\baa c_F=\lim_{\e\to 0} e^{-rT}J_\e(0,0,S(0),0).\label{clim}\eaa
\item
The value function $J=J_\e$ satisfies the Bellman equation
 \baa
&&J_t+  \max_{u\in [d_0,d_1]}  \{J_x' \f_\e + J_y' h_\e\}+ J_s'
rs+{\scriptstyle \frac{1}{2}} J_{ss}''\s^2s^2 = 0,\nonumber
\\
&&J(x,y,s,T)=\gg_\e(x,y). \label{BelEq0}\eaa The Bellman equation
has unique solution in the class of functions $J=J_\e(x,y,s,t)$ such
that $J_\e(x,y,s,t)=V_\e(x,y,\log s,t)$ for some function
$V_\e\in\X_2$. The Bellman equation  holds  as an equality that is
satisfied for a.e.
$(x,y,s,t)\in\R\times\R\times(0,+\infty)\times[0,T]$.
\end{enumerate}
\end{theorem}

\section*{Appendix:
Proofs}\setcounter{equation}{0}
%{\Large \bf Appendix: Proof of Theorems}
\renewcommand{\theequation}{A.\arabic{equation}}
%\csname @addtoreset\endcsname{equation}{section}
\renewcommand{\thelemma}{A.\arabic{lemma}}
%\csname @addtoreset\endcsname{equation}{section}
\renewcommand{\theproposition}{A.\arabic{proposition}}
\par
{\it Proof of Theorem \ref{ThM}.} It is known that the discounted
wealth is a martingale under $\P_*$ for any admissible strategies.
For a given $u(\cdot)$, the ability to fulfill the option
obligations means that $ X(T)\ge F_u$ for any $u(\cdot)$ i.e.,  $\ww
X(T)\ge e^{-rT}F_u$. Hence $X(0)=\E_*\ww X(T)\ge e^{-rT}F_u$. It
follows that \baaa c_F\ge e^{-rT}\sup_{u\in \U}\E_* F_u. \eaaa

Further, suppose that there exists $\e>0$ such that, for all
$u(\cdot)\in\U$,
 \baaa c_F\ge
e^{-rT}\E_* F_u+\e. \eaaa In this case, for any strategy
$u(\cdot)\in\U$, the claim $F_u$ can be replicated with the initial
wealth $X_0\le c_F-\e$. Therefore, any potential option buyer could
save $\e>0$ quantity of cash if she select to replicate the payoff
$F_u$ with some self-financing strategy. Therefore, a rational
option buyer would't buy an option for the price $c_F$. This
completes the proof. $\Box$
\par
{\it Proof of Lemma \ref{lemmaEc}.}
 Let $\H$ be the Hilbert space formed as the completion of the set  of all square
integrable and adapted processes in the norm of
$L_2([0,T]\times\O)$. The set $\U_1$ is a convex and closed (and,
therefore, weakly closed) subset of $\H$. Hence $\U_1$ is compact in
the weak topology of $\H$. \index{The remaining part of the proof is
similar to the proof of Lemma \ref{lemmaE}.} Consider the mapping
$\phi:\U_1\to\R$ such that $\phi(u)=\E_*F_{u}$. Let $\{u_j\}\subset
\H$ be a sequence such that \baa \phi(u_j)\to
\sup_{u\in\H}\phi(u)\quad \hbox{as}\quad
j\to+\infty.\label{phi11c}\eaa There exists a subsequence $\{u_k\}$
and $\oo u\in\U_1$ such that $u_k\to \oo u\in\H$ weakly in $\H$ as
$k\to+\infty$. By Mazur's Theorem (Theorem 5.1.2 from Yosida
(1995)), there exists a sequence of integer numbers $k=k_i\to
+\infty$ such that there exist sets of real numbers
$\{a_{mk}\}_{m=1}^k\subset[0,1]$ such that $\sum_{m=1}^ka_{mk}=1$
and that \baa &&\ww u_k\defi \sum_{m=1}^ka_{mk} u_m\to \oo u\quad
\hbox{in}\quad \H. \eaa In addition, there exists a subsequence
$\{\w u_m\}$ of this sequence such that $\w u_m\to \oo u$ a.e. as
$k\to+\infty$.
\def\u{{\rm u}}
 Consider
mappings $G(\u,y):\U_1\times C([0,T]\to \R)$ and $I(\u,y):\U_1\times
C([0,T]\to \R)$ such that $I(\u,y)=\int_0^T\u(t)f(y(t),t)dt$ and
$G(\u,y)=g(I(\u,y))$. Clearly, $G(\w u_k,S(\cdot))\to G(\oo
u,S(\cdot))$ a.s. as $k\to +\infty$. In addition, \baaa
&&|I(\u,S)|\le T\max_{t\in[0,T]}|f(S(t),t)|\max \u(t)\le \const\cdot
d_1T\max_{t\in[0,T]}(S(t)+1),\\&&|G(\u,S)|\le \const
I(\u,S(\cdot)).\eaaa By the Lebesgue's Dominated Convergence
Theorem, it follows that $\phi(\w u_k)\to \phi(\oo u)$.
\par By linearity of $I$, we have that $
  I(\w
u_k,S(\cdot))= \sum_{m=1}^ka_{mk} I(u_m,S(\cdot)).$  By the
concavity of $g$, it follows that \baaa  g(I(\w u_k,S(\cdot)))\ge
g\left(\sum_{m=1}^ka_{mk} I(u_m,S(\cdot))\right).\eaaa Hence \baaa
\phi(\w u_k)\ge \sum_{m=1}^ka_{mk} \phi(u_m)\to
\sup_{u\in\U_1}\phi(u)\quad \hbox{as}\quad k\to +\infty. \eaaa Hence
$\phi(\oo u)=\sup_{u\in\H}\phi(u)$, i.e., $\oo u$ is a optimal
control. $\Box$
\par
 {\it Proof of Theorem \ref{ThBEc}.} Let us prove statement (i).
 By Lemma \ref{lemmaEc},
 $c_F=e^{-rT}\E_*F_{\w u}g$ for some $\w u\in\U_1$. Let $\w y(t)=\int_0^t\w
 u(s)ds$.
 By the assumptions on $\ggg_\e,\phi_\e$, it follows that $e^{-rT}J_\e(0,0,S(0),0)\le c_F$ for any
 $\e>0$ and
\baaa J_\e(0,0,S(0),0)\ge \E_*\ggg_\e\left(\int_0^T \ff_\e(\w
u(t),\w y(t),S(t),t)dt\right). \eaaa Moreover, \baaa
\ggg_\e\left(\int_0^T \ff_\e(\w u(t),\w y(t),S(t),t)dt
 \right)\to
 g\left(\int_0^T\w u(t)\Ind_{\{\w y(t)<1
\}} f(S(T),t)dt\right)=F_{\w u}\quad \hbox{a.s. as}\quad
 \e\to 0. \eaaa  By the Lebesgue's Dominated Convergence
Theorem and by the assumptions on $\ggg_\e,\phi_\e$, statement (i)
follows.
\par
Let us prove statement (ii). Let us consider the change of variables
$R(t)=\ln S(t)$. Using the Ito formula, we obtain that this change
of variables transfers the corresponding control problem as \baaa
\hbox{Maximize}\quad &&\E_*\ggg_\e(x(T)) \quad\hbox{over}\quad
u(\cdot)\in \U,\label{AsOlogc}\\
 \hbox{subject to}\quad && d x(t)=\ff_\e(u(t),y(t),e^{R(t)},t)dt,
\nonumber\\
&&d y(t)=u(t)dt,
\nonumber\\
&&dR(t)=(r -\s^2/2)dt+\s dw_*(t). \label{eqxc}\eaaa
 Consider the
corresponding value function \baaa
 V(x,y,z,t)\defi
\sup_{u(\cdot)\in\U}\E_*\Bigl\{\ggg_\e\left(x(T)\right)\Bigl|
x(t)=x,\, y(t)=y,\,R(t)=z\Bigr\}. \label{JVc}\eaaa Again, the
coefficients of this  problem are such that the assumptions of
Theorem 4.1.4 and Theorem 4.4.3 from Krylov (1980), p.167,192, are
satisfied. \par
 By Theorem
4.1.4 from Krylov (1980), p.167, this function satisfy the
corresponding parabolic Bellman equation that has unique solution
$V\in\X_1$. The Bellman equation  holds in the generalized sense,
i.e., as an equality of the distributions. This equation includes
only one partial derivative of the second order, $V''_{zz}$
presented with the coefficient $\s^2/2>0$. By Theorem 4.4.3 from
Krylov (1980), p.192, the derivative $V'_t(x,y,z,t)$ belongs to
$\X$. It follows that $V''_{zz}\in\X$. The Bellman equation for
$J(x,y,s,t)=V(x,y,\log s,t)$ is defined by the equation for $V$ with
the corresponding change of variables.
 Then the proof of statement (ii) and Theorem \ref{ThBEc} follows.  $\Box$.

 \par
{\em Proof} of Theorem \ref{ThBEl} is similar to the proof of
Theorem \ref{ThBEc}.

\par
{\it Proof of Theorem \ref{ThAM}.} First, note that, by Lemma
\ref{lemmaEc}, it follows that a optimal control $u\in \U_1$ exists.
Consider a linear mapping $\phi(u):\U_1\to\R$ such that
$\phi(u)=\E_*F_u$.
\par
Let  $u(\cdot)\neq \w u(\cdot)$ be a process in $\U_1$. Let us show
that $u(\cdot)$ cannot be optimal.  Since $u(\cdot)\neq \w
u(\cdot)$,
 there exist a  equivalent in $L_2([0,T]\times \O)$ version of $u(\cdot)$, $v\in[0,L]$,
 non-random times $t_0\in[0,T-L^{-1}]$, $t_1\in (T-L^{-1},T]$ and a set $\O_0\in\F_{t_0}$,  such that
$\P_*(\O_0)>0$,  $t_k$ are Lebesgue points for $u(t,\o)f(S(t,\o),t)$
for all $\o\in\O_0$, and, for small  enough $\e>0$,
 \baaa
 &&u(t_0,\o)> v, \quad V(t,\o)\defi u(t,\o)+u(t-t_1+t_0,\o)-v\in [0,L]\quad
  \forall t\in J_\e,\ \o\in\O_0, \eaaa where  $J_\e\defi[t_1-\e,t_1)$.
  Note that feasibility of the property that $V(t,\o)\in[0,L]$ can be seen from the existence of
  $t_0,t_1,\O_0$ such that  $\esssup_{t\in I_{\e}}(u(t,\o)-v)<0$ and $\esssup_{t\in J_{\e}}u(t,\o)<L$ for all
  $\o\in\O_0$.
\par
   Let  $I_\e\defi [t_0,t_0+\e)$ and let $u_\e(t)$ be constructed as the following: \baaa
u_\e(t)=\begin{cases}u(t),\quad t\notin I_\e\cup  J_\e,\\
v,\quad t\in I_\e,\\  V(t),\quad t\in
 J_\e.\end{cases}
 \eaaa
Let us show that  $u_\e(\cdot)\in\U_1$ for  small enough $\e>0$.
Clearly, this process is adapted to $\F_t$. By the definition of
$\U_1$, \baaa
\int_0^Tu(t)dt=\int_0^{t_0}u(t)dt+\int_{t_1}^Tu(t)dt+\int_{I_\e\cup
J_\e}u(t)dt=1 \eaaa and \baaa
&&\int_0^Tu_\e(t)dt=\int_0^{t_0}u(t)dt+\int_{t_1}^Tu(t)dt+\e
v+\int_{J_\e}V(t)dt\\&&= \int_0^{t_0}u(t)dt+\int_{t_1}^Tu(t)dt+\e
v+\int_{J_\e}u(t)dt+\int_{J_\e}u(t-(t_1-t_0))dt-\e
v\\&&=\int_0^Tu(t)dt=1,
 \eaaa
 since $\int_{J_\e}u(t-t_1+t_0)dt=\int_{J_\e}u(t-t_1+t_0)dt=1$.
 Hence $u_\e(\cdot)\in\U_1$ for  small enough $\e>0$.
 \par
 To prove that
$u(\cdot)$ is not optimal, it suffices to show that \baa \lim_{\e\to
0+}\frac{\phi(u_\e)-\phi(u)}{\e}>0. \label{lim1}\eaa
\par
 We have that \baa &&\lim_{\e\to
0+}\frac{\phi(u_\e)-\phi(u)}{\e}\nonumber\\&&=\lim_{\e\to
0+}\e^{-1}\E_*\Ind_{\O_0}\left[\int_{I_\e}
(v-u(t))f(S(t),t)dt+\int_{J_\e} (V(t)-u(t))f(S(t),t)dt
\right]\nonumber\\&&=\lim_{\e\to
0+}\E_*\Ind_{\O_0}\left[\e^{-1}\int_{I_\e}
(v-u(t))f(S(t),t)dt+\e^{-1}\int_{J_\e} (V(t)-u(t))f(S(t),t)dt
\right]\nonumber\\&&=\E_*\Ind_{\O_0}\left[(v-u(t_0))f(S(t_0),t_0)+
(u(t_1)+u(t_0)-v-u(t_1))f(S(t_1),t_1)
\right]\nonumber\\&&=\E_*\Ind_{\O_0}\left[(v-u(t_0))f(S(t_0),t_0)+(
u(t_0)-v)f(S(t_1),t_1)
\right]\nonumber\\&&=\E_*\Ind_{\O_0}\E_*\left\{(v-u(t_0))f(S(t_0),t_0)+(
u(t_0)-v)f(S(t_1),t_1)|\F_{t_0} \right\}\nonumber\\
&&=\E_*\Ind_{\O_0}(u(t_0)-v) \left[\E_*\left\{f(S(t_1),t_1)|\F_{t_0}
\right\}-f(S(t_0),t_0)\right].
 \label{lim2}\eaa
 \par
 Further, we have that $\ww S(t)\defi S(t)e^{-rt}$ is  a martingale under
$\P_*$. We have also that the support of the conditional
distribution of  $S(t_1)$ given $S(t_0)$ is $(0,+\infty)$. Since
$g(\cdot)$ is convex and non-linear, it follows from the Jensen's
inequality that
 \baaa
\E_*\{h(S(t_1))|F_{t_0}\}=\E_*\{h(e^{rt_1}\ww S(t_1))|F_{t_0}\}>
h(e^{rt_1}\ww S(t_0))=h(e^{r(t_1-t_0)} S(t_0)). \eaaa

By the properties of $h$, we have that \baaa
e^{-r(t_1-t_0)}h(e^{r(t_1-t_0)} S(t_0))\ge h(S(t_0)). \eaaa  Hence
\baaa h(e^{r(t_1-t_0)} S(t_0))\ge e^{r(t_1-t_0)}h(S(t_0)), \eaaa and
\baaa \E_*\{h(S(t_1))|F_{t_0}\}> h(e^{r(t_1-t_0)} S(t_0))\ge
e^{r(t_1-t_0)}h(S(t_0)). \eaaa  Hence \baaa
\E_*\{f(S(t_1),t_1)|F_{t_0}\}=e^{r(T-t_1)}\E_*\{h(S(t_1))|F_{t_0}\}>
e^{r(T-t_1)} e^{r(t_1-t_0)}h(S(t_0))=f(S(t_0),t_0). \eaaa By
(\ref{lim2}), we obtain that limit (\ref{lim2}) is positive. Hence
(\ref{lim1}) holds and the proof follows. $\Box$

\par
 {\it Proof of Theorem \ref{ThBE}.} Let us prove statement (i). Let $\{u_i\}\subset\U$ be a sequence such
that \baa \E_*F_{u_i}\to \sup_{u\in\U}\E_*F_{u}\quad \hbox{as}\quad
i\to+\infty.\label{phi1c}\eaa
 \index{It suffices to consider the
case when $d_0=0$ only, since if $d_0>0$ then (\ref{clim}) holds
since  $J_\e(0,0,S(0),0)=c_F$ for any $\e>0$.  Let  $u_{i\e}(t)=
 u_i(t)$, $t<T-\e$, and let $u_{i\e}(t)=d_1$, $t>T-\e$.} Let $F_u'$ be defined similarly to $F_u$ with $(f,g)$ replaced by
$(\phi_\e,g_\e)$. Clearly, for any $u\in\U$, there exists $\w
u\in\U$ such that $\E_* F'_{\w u}={\cal J}_\e(u,0,0,S(0),0)$. Hence
$J_\e(0,0,S(0),0)\le \sup_{u\in\U}\E_* F'_{u}$. By the properties of
$(\f_\e,\gg_\e)$, it follows that $\sup_{u\in\U}\E_* F'_{u}\le
e^{rT}c_F $. Hence $e^{-rT}J_\e(0,0,S(0),0)\le c_F$ for any
 $\e>0$. Moreover,
\baaa J_\e(0,0,S(0),0)\ge \E_*\gg_\e\left(\int_0^T
\f_\e(u_{i}(t),S(t),t)dt,\int_0^T h_\e (u_i(t),t)dt \right). \eaaa
\par
Let $i$ be fixed.
 If $\int_0^Tu_i(t)dt>0$ then, by the Lebesgue's Dominated Convergence
Theorem and by the assumptions on $g,f$, \baaa \gg_\e\left(\int_0^T
\f_\e(u_{i}(t),S(t),t)dt,\int_0^T h_\e(u_{i}(t),t)dt
 \right)\to
 g\left(\frac{\int_0^Tu_i(t)f(S(T),t)dt}{\int_0^Tu_i(t)dt }\right)=F_{u_i}\quad \\
 \hbox{a.s. as}\quad
 \e\to 0. \eaaa
 \par
If $\int_0^Tu_i(t)dt=0$ then, by assumption (\ref{delta}), \baaa
\gg_\e\left(\int_0^T \f_\e(u_{i}(t),S(t),t)dt,\int_0^T
h_\e(u_{i}(t),t)dt
 \right)\\
=g_\e\biggl(\int_{T-\e}^{T-\e+\e^2}
h_\e(u_i(t),t)f(S(t),t)dt+\int_{T-\e+\e^2}^T d_1f(S(t),t)dt,\\
 \int_{T-\e}^{T-\e+\e^2} h_\e(u_i(t),t)dt+\int_{T-\e+\e^2}^T
d_1dt\biggr)
\\
=g_\e\left(O(\e^2)+\int_{T-\e+\e^2}^T
d_1f(S(t),t)dt,O(\e^2)+(\e-\e^2)d_1\right)
\\
\to g(f(S(T),T))=F_{u_i}
 \quad \hbox{a.s. as}\quad
 \e\to 0 \eaaa
 again. By the Lebesgue's Dominated Convergence
Theorem and by the assumptions on $g,f$ again, it follows that \baa
{\cal J}_\e(u_i,0,0,S(0),0)\to \E_*F_{u_i}\quad\hbox{as}\quad \e\to
0.\label{Jlim} \eaa We now in the position to prove statement (i).
It suffices to show that, for any $\d>0$, there exists $\e_*>0$ such
that $J_\e(0,0,S(0),0)\ge \sup_{u\in\U}\E_*F_{u}-\d$ for $\e<\e_*$.
Let $i$ be such that  $i$ such that $\E_*F_{u_i}\ge
\sup_{u\in\U}\E_*F_{u}-\d/2$. By (\ref{Jlim}),  there exists
$\e_*=e_*(\d,i)>0$ such that ${\cal J}_\e(u_i,0,0,S(0),0)\ge
\E_*F_{u_i}-\d/2$ for all $\e<\e_1$. Hence ${ J}_\e(0,0,S(0),0)\ge
\E_*F_{u_i}-\d/2$ and ${ J}_\e(0,0,S(0),0)\ge
\sup_{u\in\U}\E_*F_{u}$ for these $\e$. \index{ Hence for any
$i,\d>0$, there exists $\e=\e(i,\d)>0$ such that ${\cal
J}_\e(u_i,0,0,S(0),0)\ge \E_*F_{u_i}-\d$. Hence ${
J}_\e(0,0,S(0),0)\ge \E_*F_{u_i}-\d$ for this $\e$.} Then statement
(i) follows.
%\end{document}
\par
Let us prove statement (ii). Let us consider the change of variables
$R(t)=\ln S(t)$. Using the Ito formula, we obtain that this change
of variables transfers the corresponding control problem as \baa
\hbox{Maximize}\quad &&\E_*\gg_\e(x(T),y(T)) \quad\hbox{over}\quad
u(\cdot)\in \U,\label{AsOlog}\\
 \hbox{subject to}\quad && d x(t)=\f_\e(u(t),e^{R(t)},t)dt,
\nonumber\\
&&d y(t)=h_\e(u(t),t)dt,
\nonumber\\
&&dR(t)=(r -\s^2/2)dt+\s dw_*(t). \label{eqx}\eaa
 Consider the
corresponding value function \baa
 V(x,y,z,t)\defi
\sup_{u(\cdot)\in\U}\E_*\Bigl\{\gg_\e\left(x(T),y(T)\right)\Bigl|
x(t)=x,\, y(t)=y,\,R(t)=z\Bigr\}. \label{JV}\eaa Note that the
coefficients of this  problem are such that the assumptions of
Theorem 4.1.4 and Theorem 4.4.3 from Krylov (1980), p.167,192, are
satisfied. The remaining part of the proof repeats the proof of
Theorem \ref{ThBEc}(ii). This completes the proof of Theorem
\ref{ThBE}. $\Box$

\section*{References} $\hphantom{xx}$
 Bender, C., and Schoenmakers, J. (2006). An iterative method
for multiple stopping: convergence and stability. {\it Advances in
Applied Probability} {\bf 38}, no. 3, 729–749.

Bender, C. (2011). Dual pricing of multi-exercise options under
volume constraints. {\em Finance and Stochastics} {\bf 15}, Iss. 1,
1--26.

Briys, E., Mai, Huu Minh, Bellalah, M., de Varenne, M.. (1998).
Options, Futures and Exotic Derivatives.al. Chichester, New York
Wiley.

Carmona, R. and Dayanik, S. (2006). Optimal multiple stopping of
linear diffusions and swing options. Princeton University. (tech.
report).

Carmona, R., Touzi, N. (2006). Optimal multiple stopping and
valuation of swing options. {\it Mathematical Finance}, forthcoming.

Dai, Min Kwok, Yue Kuen (2008). Optimal multiple stopping models of
reload options and shout options.  {\it Journal of Economic Dynamics
and Control} {\bf 32},  Iss. 7, 2269--2290.

Delbaen, F., and Yor, M. (2002). Passport options. {\em Mathematical
Finance} {\bf 4} Iss. 4, 299--328.

Dokuchaev, N.G. (2009).  Multiple rescindable options and their
pricing.   {\it International Journal of Theoretical and Applied
Finance (IJTAF)}, {\bf 12}, Iss. 4, 545 - 575.

Hyer, T., Lipton-Lifschitz, A. and Pugachevsky, D. (1997). Passport
to Success, Risk 10 (1997), September, 127–131.

Kampen, J. (2008). Optimal Strategies of Passport Options. {\em
Mathematics in Industry} {\bf 12}, 666-670.

 Kifer, Yu. (2000). Game options, {\it Finance and Stochastics} {\it 4},
443-463.

Kyprianou, A.E. (2004). Some calculations for Israeli options, {\it
Finance and Stochastics}  {\it 8}, 73-86.

Kramkov, D.O. and Mordecky, E. (1994), An integral option, {\it
Theory of Probability and Its Applications} {\bf 1}, 162-172.

\par
Krylov, N.V.  (1980).  {\em Controlled Diffusion Processes}.
Springer-Verlag, New York.

Meinshausen, N., Hambly, B. M. (2004). Monte Carlo Methods for the
valuation of multiple-exercise options. {\it Mathematical Finance}
{\bf 14}, 557-583.

Nagayama, I. (1999). Pricing of Passport Option, {\em J. Math.
Sci.Univ. Tokyo } {\bf 5} 747-785.

Peskir, G. (2005). The Russian option: finite horizon, {\it Finance
and Stochastics}  {\bf 9}, iss. 2, 251-267.

 Villeneuve, S. (1999). Exercise regions of American options on several assets,
{\it Finance and Stochastics} {\bf 3}, 295-322.

Yosida, K. (1995) {\it Functional Analysis.}  Springer, Berlin
Heilderberg New York.

\end{document}